\documentclass{llncs}
\usepackage[T1]{fontenc}
\usepackage[colorlinks=true, linkcolor=blue, citecolor=blue, urlcolor=blue]{hyperref}
\usepackage{graphicx}
\usepackage{booktabs}
\usepackage{tabularx}
\usepackage{placeins}
\usepackage{amsmath}
\usepackage{orcidlink}
\usepackage{fancyhdr}

\begin{document}
\title{To Use or Not to Use: Investigating Student Perceptions of Faculty Generative AI Usage in Higher Education}
\titlerunning{Student Perceptions on Faculty GenAI Use}
%
%
\author{Jie Gao\inst{1,3}\orcidlink{0000-0002-6933-950X} \and
Jiayi Zhang\inst{2}\orcidlink{0000-0002-7334-4256} \and
Dan Chen\inst{4}}
\authorrunning{J. Gao et al.}
%

\institute{McGill University, Montreal, Canada\\
University of Pennsylvania, Philadelphia, USA\\
Mila – Quebec Artificial Intelligence Institute, Montreal, Canada \\
University of Toronto, Toronto, Canada\\
\email{{jie.gao3}@mail.mcgill.ca\\{joycez}@upenn.edu\\
{en14dc}@gmail.com}}
\maketitle              
\begin{abstract}
While Generative AI (GenAI) rapidly integrated into higher education, existing research has primarily focused on regulating student use. As a result, student perspectives on faculty adoption of GenAI remained unexplored. In this study, we analyzed survey responses from 156 undergraduate and graduate students to examine their attitudes toward both student and faculty use of GenAI. We classified students into four groups based on their attitudes, including GenAI Optimists, Student Support Group, Faculty Support Group, and Non-supporters. Findings show that 37\% of participants do not support GenAI use by either students or faculty, while 31\% support GenAI use in both contexts. 
We also conducted thematic analysis to understand participants' concerns on faculty GenAI usage. Results revealed that (1) a majority of students (79\%) questioned the validity and reliability of GenAI-generated responses, and (2) 37\% of students feared that faculty overreliance on GenAI created a “futile cycle” that might reduce faculty critical thinking. Our findings showed that students expressed concerns about GenAI use by faculty in teaching and grading contexts, with pedagogical concerns being most prominent. These findings informed the future use of GenAI in teaching and learning in higher education.

\end{abstract}

\section{Introduction}
The rapid advancement and public availability of generative artificial intelligence (GenAI) tools, such as ChatGPT, have fundamentally changed the landscape of higher education. These technologies are increasingly used by students to support learning activities, including brainstorming, explaining concepts, and receiving formative feedback~\cite{Kasneci2023ChatGPTGood,Zhai2022ChatGPTUX,Arowosegbe2024GenAIUK}. At the same time, instructors are beginning to explore GenAI for instructional design, teaching support, and assessment-related tasks, such as generating practice problems, providing feedback, or assisting with grading~\cite{Baidoo2023EducationGenAI,Barus2025AIGovernance}. As GenAI becomes embedded across educational practices, understanding how different stakeholders perceive its appropriate use has become a pressing concern for institutions of higher education.

Existing research has documented growing student adoption of GenAI tools for learning purposes, often emphasizing perceived benefits such as efficiency, accessibility, and personalized support~\cite{Barus2025AIGovernance,Arowosegbe2024GenAIUK}. However, the integration of GenAI into faculty teaching and assessment practices raises distinct pedagogical, ethical, and trust-related questions~\cite{Barus2025AIGovernance}. While faculty use of GenAI may offer opportunities for scalability and instructional support, it also challenges long-standing assumptions about instructional responsibility, assessment fairness, and the human role in teaching~\cite{Barus2025AIGovernance}. Importantly, students may evaluate GenAI use differently depending on whether it is used by themselves or by instructors, leading to potential asymmetries in acceptance and perceived legitimacy.

Emerging evidence suggest a tension in which students may view their own use of GenAI as a legitimate learning aid, while expressing discomfort or resistance toward faculty use of the same technologies for teaching or grading~\cite{Chan2024ReplaceTeachers}. Such asymmetries, if present, have significant implications for instructional design, assessment transparency, and institutional policy.

To address this gap, the present study examines students’ comparative attitudes toward GenAI integration in higher education, distinguishing between GenAI use for student learning and for faculty teaching and assessment. The study focuses on identifying whether an asymmetry exists in students’ acceptance of GenAI use across these contexts and on unpacking the specific concerns that drive student resistance to faculty use of GenAI, particularly in teaching and grading. By foregrounding student perspectives, this work aims to provide empirically grounded insights into how GenAI can be integrated in ways that are pedagogically sound, ethically responsible, and aligned with student expectations.

Specifically, this study addresses the following research questions:

\textbf{RQ1.} What are students’ perceptions of students using GenAI in learning and faculty using GenAI in teaching and grading?

\textbf{RQ2.} What are the primary concerns driving student resistance to the integration of GenAI into teaching and grading by faculty?

\section{Related Work}
The integration of GenAI in education has been increasingly prominent, as it provides personalized learning experiences~\cite{marrone2025understanding}, automates academic feedback~\cite{lu2025exploring}, simplifies pedagogical preparation~\cite{tillmanns2025mapping}, and identifies students' learning abilities~\cite{huang2025ssrlbot}.
A growing body of research has evaluated the performance of GenAI in higher education~\cite{aithal2023effects,belkina2025implementing,huang2025ssrlbot}. However, students’ perspectives on GenAI vary. A majority of research on student perceptions of GenAI indicates both positive and passive attitudes toward using it for academic purposes~\cite{bhullar2024chatgpt,chan2023students}. Recent research~\cite{marrone2025understanding} on student perceptions of GenAI as a teammate shows that most students perceive it as a useful tool, but remain concerns about its capabilities and trustworthiness. Additionally, academic integrity issues, such as plagiarism, are a significant concern in the use of generative AI in higher education~\cite{bhullar2024chatgpt,gao2024can}. However, limited research investigates students’ perspectives on faculty use of GenAI, although GenAI has been widely used as a teaching assistant and evaluation tool.


\section{Methodology}

\subsection{Dataset and Measurement}

We drew an open-access dataset on student perceptions of both student and faculty use of GenAI for coursework~\cite{DVN_B2EAUL_2025} hosted on the Harvard Dataverse. 
The data were collected through an anonymous online survey administered via Google Forms. The survey includes demographic information (e.g., university, major, minor, and year of study), 11 items on students' perceptions of their own use of GenAI for coursework, and 5 items on students' perceptions of faculty use of GenAI for the same coursework. The present analysis selects four related questions: (1) \textit{``I think faculty should/should not use generative AI in their teaching.''} (2) \textit{``I think faculty should/should not use generative AI to grade student assignments.''} (3) \textit{``I should/should not use generative AI in my assignments.''} and (4) \textit{``What concerns do you have about faculty using generative AI?''}

To examine the perception of the use of GenAI in higher education, we filtered the dataset to only include  responses from undergraduate and graduate students who have completed the survey (N = 156). Participants are primarily pursuing their studies in universities based in the United States, Canada, the United Kingdom, Australia, and France. Their majors vary across Arts, Psychology, Business, Computer Science, Data Science, Engineering, Health, Medicine, Education, and Law.

The dataset was selected for three primary reasons: First, this survey is suitable for investigating the students' attitude toward GenAI usage in the educational domain, as it explicitly juxtaposes items regarding student agency (usage for coursework) against faculty assessment (usage for grading), which is crucial for detecting the asymmetries. Second, it offers authentic responses from a diverse higher education cohort, including both undergraduate and graduate levels. Third, it includes open-ended textual data grounded in immediate student concerns.

\vspace{1em}
\begin{table}[t!]
\centering
\caption{Classifying the Open-ended responses into four concern themes.}
\label{t1}

\begin{tabularx}{\linewidth}{
  >{\raggedright\arraybackslash}p{0.2\linewidth}
  >{\raggedright\arraybackslash}X
  >{\raggedright\arraybackslash}X
}
\toprule
\textbf{Concern Theme} & \textbf{Description} & \textbf{Examples} \\
\midrule

Professional Concern
&
Students think that faculty’s reliance on GenAI may lead to laziness and a loss of essential engagement.
&
\textit{``I am worried it will make teacher too lazy.''}\newline
\textit{``Faculty could be less involved in teaching.''} \\
\midrule

GenAI Quality Concern
&
Students think that GenAI has the potential for hallucinations and issues with accuracy, validity, and reliability.
&
\textit{``Errors in information and grading.''}\newline
\textit{``That they would grade our assignments wrong or would give us false information.''} \\
\midrule

Pedagogical Concern
&
Students think that faculty’s long-term overreliance on GenAI may reduce their critical thinking skills and teaching quality.
&
\textit{``I am thinking AI usage may distance the relationship between faculty in students in that faculty are less aware of where students stand or learn.''}\newline
\textit{``My concern is that teaching methods will lack individuality.''} \\
\midrule

Ethical Concern
&
Students think that GenAI may cause ethical issues, such as plagiarism and bias.
&
\textit{``I'm mainly worried about the ethical and environmental concerns that surround certain AI technologies.''}\newline
\textit{``Academic integrity issues, such as students using AI to complete assignments or cheat on exams.''} \\

\bottomrule
\end{tabularx}
\end{table}

\subsection{Data Processing and Analysis}
To explore whether students hold asymmetric views about their own and faculty use of GenAI in completing and grading assignments, we conducted a McNemar’s test to assess whether students’ attitudes were consistent across these two contexts, based on students' responses to Questions 2 ( \textit{``I think faculty should/should not use generative AI to grade student assignments.''}) and 3 ( \textit{``I should/should not use generative AI in my assignments.''}).

To further understand students' positions on GenAI use, based on their responses to the two questions, we grouped students into four clusters based on their attitudes toward GenAI usage. 1) GenAI Optimists group represent students who support GenAI integration for both students and faculty. 2) Student Support group represents students who support their own use of GenAI but oppose faculty use for grading. 3) Faculty Support group represents students who oppose student use of GenAI but accept faculty use for grading. 4) Non-supporters represent students who do not support GenAI use in either context.

Furthermore, to explore students’ concerns about faculty use of GenAI for coursework, we extracted participants’ responses to the open-ended question \textit{``What concerns do you have about faculty using generative AI?''} and conducted thematic analysis, a method that identifies, organizes, and provides insights into themes through analyzing qualitative data \cite{kushnir2025thematic}. In total, 130 participants (83.3\%) shared their concerns to this question. 

Guided by previous research~\cite{gao2026investigatingselfregulatedlearningsequences}, we  developed a thematic coding scheme to analyze students’ responses to the concern question on GenAI use. Table~\ref{t1} outlines the four themes of concern and their corresponding descriptions and examples, which include Professional Concern, GenAI Quality Concern, Pedagogical Concern, and Ethical Concern.
The responses were classified into four categories and these categories are not mutually exclusive, meaning that a response may address more than one concern. We removed duplicate or non-substantive responses (e.g., "none", "no concerns"). We coded all responses and compared the results, achieving an agreement of $\sim$~92\%. Inconsistently coded data were reviewed and resolved through discussion.

\section{Results}

\subsection{Students' Perceptions of GenAI Use (RQ1)}
To investigate whether students perceive an asymmetry in GenAI usage between themselves and faculty, we perform a McNemar’s test on paired nominal data that compares student acceptance of their own use of GenAI in assignments with their acceptance of faculty use of GenAI for grading.

    

\begin{figure}[t!]
    \centering
    \includegraphics[width=\columnwidth]{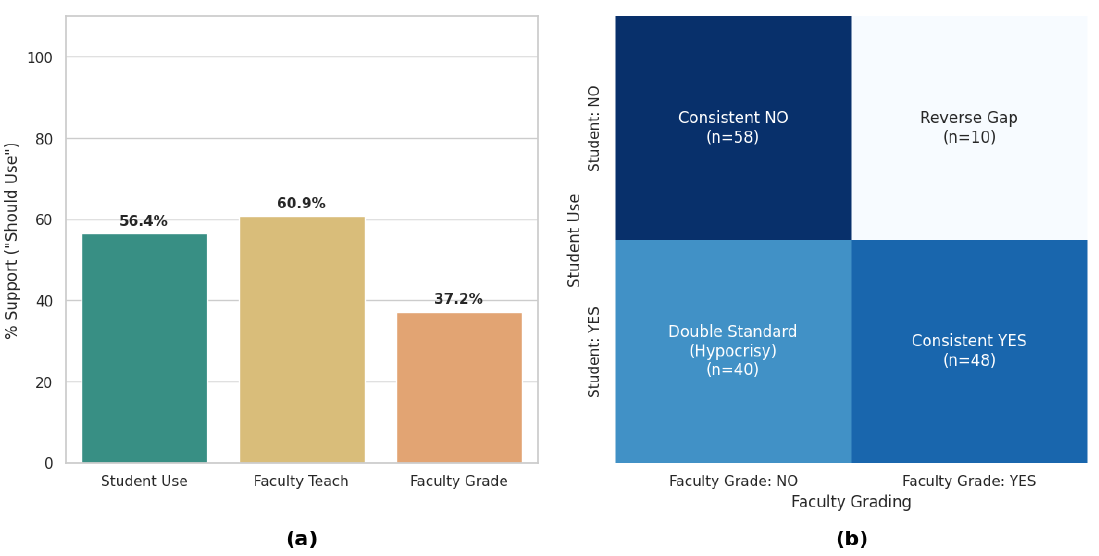}
    \caption{Student Attitudes Toward Student Use of GenAI and Faculty Use of GenAI.}
    \vspace{0pt}
\label{f1}
\end{figure}

Our results indicate a statistically significant divergence in student attitudes
($\chi^2 = 16.820, p < .001$). As shown in Figure~\ref{f1}a, while 56.4\% of
students ($N = 88$) believe they should be allowed to use GenAI in assignments,
only 37.2\% ($N = 58$) support faculty use of GenAI for grading those assignments. 

The cross-tabulation reveals four distinct clusters of student perceptions of GenAI use (see Figure~\ref{f1}b). A majority of students (37.2\%, $N = 58$), as \textbf{Non-supporters}, did not support GenAI use by either students or faculty for coursework. In contrast, 30.8\% of students (\textbf{GenAI Optimists}) supported GenAI use in both contexts. Notably, 25.6\% of students (\textbf{Student Support Group}) preferred student use of GenAI but not faculty use, indicating an asymmetry in their perceptions between students and faculty. Only a minority of students (\textbf{Faculty Support Group}), who opposed student use of GenAI but supported faculty use.


\subsection{Primary Concerns on Faculty GenAI Use (RQ2)}
\begin{figure}[t!]
    \centering
    \includegraphics[width=0.8\columnwidth]{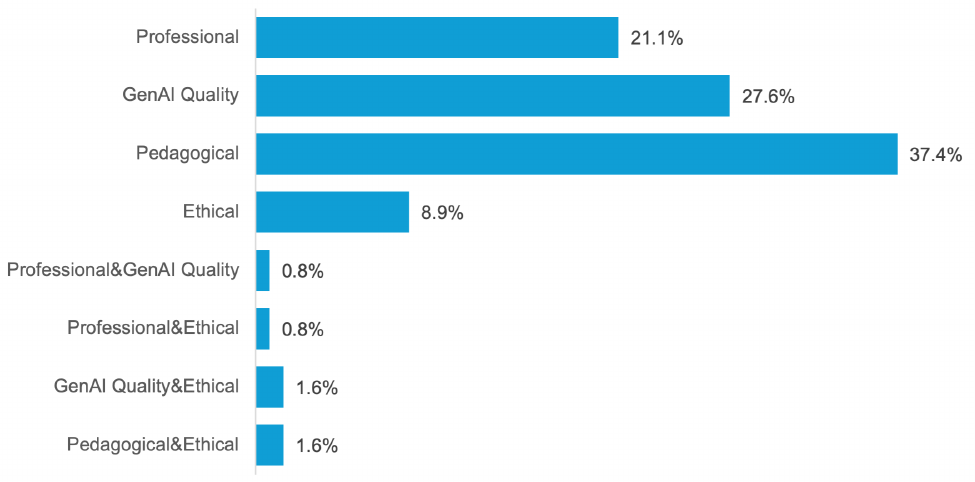}
    \caption{Distribution of concern categories.}
    \vspace{0pt}
\label{f4}
\end{figure}

To understand the underlying reasons for student resistance to faculty use of GenAI in teaching and grading, we conducted a thematic analysis of the open-ended responses. Among 156 participants, 123 provided valid responses. As shown in Figure~\ref{f4}, the analysis reveals the distribution of four distinct concern themes: professional, GenAI quality, pedagogical, and ethical.

\textbf{Theme 1 (Professional Concern):}
Students perceived faculty use of GenAI as a form of professional dereliction. Approximately one-fifth of students (N = 26, 21\%) viewed GenAI not as a teaching assistance tool but as an abdication of instructional responsibility. Students frequently described this behavior using terms such as \textit{“lazy,” “lack of genuine passion,”} and \textit{“less involvement.”} They also reported that this behavior may weaken the connection between students and faculty.

\textbf{Theme 2 (GenAI Quality Concern):}
Students questioned the validity and reliability of GenAI, expressing concerns about its outputs. A total of 34 students reported concerns in the application of GenAI in teaching and assessment contexts. Many of them (N = 27, 79\%) believe that GenAI still exhibits accuracy issues, and they cite risks such as \textit{“inaccurate,” “unchecked,” “misuse,”} and \textit{“hallucinations”} in grading.

\textbf{Theme 3 (Pedagogical Concern):}
Students prioritized pedagogical concerns, fearing that faculty reliance on GenAI may have potentials to erode their critical thinking skills. This is the most prevalent theme (N = 46, 37\%), and students shifted their focus from grading to long-term consequences for learning. Students reported that overreliance on GenAI may inhibit learning, stunt creativity and originality, decrease the depth of teaching content, and reduce critical thinking and human interaction in the learning process.

\textbf{Theme 4 (Ethical Concern):}
Students raised critical ethical concerns regarding academic integrity, bias, and data privacy. Eleven students only reported ethical concerns and five reported multi-dimensional concerns that include this theme. Academic integrity (N = 7), data privacy (N = 4), and bias (N = 3) emerged as the main issues. For example, one student noted that GenAI use may cause \textit{"the possibility for plagiarism, mistakes, contradictory information, and a disconnect between teachers, their subject, and their students."}

Furthermore, a small subset of students expressed more than one concern. For instance, one respondent showed both GenAI quality and ethical concerns, arguing that \textit{"inaccuracies and biases in AI outputs may misinform students if not properly vetted."}



\section{Discussion and Conclusion}

This study contributes to the growing literature on GenAI in higher education by shifting the focus from student regulation to students’ perceptions of faculty adoption of GenAI. Our findings reveal that more than one-third of students oppose GenAI usage by both students and faculty. In contrast, another one-third of students support its adoption in both contexts. This polarization suggests that a significant part of the students remains apprehensive regarding the integration of GenAI across learning, teaching, and grading.

The four identified student groups (GenAI Optimists, Student Support Group, Faculty Support Group, and Non-supporters) highlight the heterogeneity in student attitudes and suggest that student perceptions cannot be captured by a simple pro- or anti-GenAI dichotomy. Across groups, students’ concerns were largely pedagogical and ethical rather than technical. Students feared that faculty overreliance on GenAI could undermine instructional quality, reduce meaningful human interaction, and weaken faculty’s critical thinking. These concerns align with broader debates about automation in professional work, where efficiency gains may come at the cost of expertise, judgment, and relational engagement.

We note several limitations. First, the analysis was conducted on a subset of survey responses, which may limit the scope of our findings and overlook other important insights into students’ perceptions of GenAI in higher education. Relatedly, the primary focus of this study was to examine the asymmetry in perceptions, rather than to investigate the underlying reasons for why this asymmetry exists. These reasons may be rooted in broader institutional, social, or individual factors that were not captured in this study. Therefore, future work should explore the interplay between student motivations and their concerns about faculty adoption of GenAI to provide a more holistic understanding.

In conclusion, students are not uniformly resistant to GenAI but hold nuanced and role-sensitive views about its appropriate use. Institutions seeking to promote GenAI adoption should attend to students’ pedagogical expectations and concerns, particularly around human involvement, accountability, and fairness. Future research should further examine how these perceptions evolve over time and how different implementation strategies shape student trust and acceptance of GenAI in academic contexts.


%

\bibliographystyle{splncs04}
\bibliography{references}

\end{document}